\documentclass[12pt]{article}
\usepackage{fancyhdr}

\usepackage{mathrsfs}
\usepackage[T1]{fontenc}
\usepackage{mathpazo}
\usepackage{setspace}
\usepackage{amsfonts}
\usepackage{amssymb}
\usepackage{amsmath}
\usepackage{epsfig}
\usepackage{latexsym}
\usepackage{color}
\usepackage{graphicx}
\usepackage{nicefrac}
\usepackage[latin1]{inputenc}
\usepackage{slashed}
\usepackage{multirow}
\usepackage{fancybox}
\usepackage{cite}
\usepackage{comment}
\usepackage{soul}

\usepackage[all,cmtip]{xy}
\usepackage{hyperref}

\def\hybrid{\topmargin -20pt    \oddsidemargin 0pt
        \headheight 0pt \headsep 0pt
        \textwidth 6.25in       
        \textheight 9.25in       
        \marginparwidth .875in
        \parskip 5pt plus 1pt   \jot = 1.5ex}

\hybrid

\def\baselinestretch{1.2}

\catcode`\@=11

\def\marginnote#1{}
\newcount\hour
\newcount\minute
\newtoks\amorpm
\hour=\time\divide\hour by60
\minute=\time{\multiply\hour by60 \global\advance\minute by-\hour}
\edef\standardtime{{\ifnum\hour<12 \global\amorpm={am}%
        \else\global\amorpm={pm}\advance\hour by-12 \fi
        \ifnum\hour=0 \hour=12 \fi
        \number\hour:\ifnum\minute<10 0\fi\number\minute\the\amorpm}}
\edef\militarytime{\number\hour:\ifnum\minute<10 0\fi\number\minute}

\def\draftlabel#1{{\@bsphack\if@filesw {\let\thepage\relax
   \xdef\@gtempa{\write\@auxout{\string
      \newlabel{#1}{{\@currentlabel}{\thepage}}}}}\@gtempa
   \if@nobreak \ifvmode\nobreak\fi\fi\fi\@esphack}
        \gdef\@eqnlabel{#1}}
\def\@eqnlabel{}
\def\@vacuum{}
\def\draftmarginnote#1{\marginpar{\raggedright\scriptsize\tt#1}}

\def\draft{\oddsidemargin -.5truein
        \def\@oddfoot{\sl preliminary draft \hfil
        \text\thepage\hfil\sl\today\quad\militarytime}
        \let\@evenfoot\@oddfoot \overfullrule 3pt
        \let\label=\draftlabel
        \let\marginnote=\draftmarginnote
   \def\@eqnnum{(\theequation)\rlap{\kern\marginparsep\tt\@eqnlabel}%
\global\let\@eqnlabel\@vacuum}  }

\def\preprint{\twocolumn\sloppy\flushbottom\parindent 2em
        \leftmargini 2em\leftmarginv .5em\leftmarginvi .5em
        \oddsidemargin -.5in    \evensidemargin -.5in
        \columnsep .4in \footheight 0pt
        \textwidth 10.in        \topmargin  -.4in
        \headheight 12pt \topskip .4in
        \textheight 6.9in \footskip 0pt
        \def\@oddhead{\thepage\hfil\addtocounter{page}{1}\thepage}
        \let\@evenhead\@oddhead \def\@oddfoot{} \def\@evenfoot{} }

\def\numberbysection{\@addtoreset{equation}{section}
        \def\theequation{\thesection.\arabic{equation}}}

\def\underline#1{\relax\ifmmode\@@underline#1\else
        $\@@underline{\hbox{#1}}$\relax\fi}

\def\titlepage{\@restonecolfalse\if@twocolumn\@restonecoltrue\onecolumn
     \else \newpage \fi \thispagestyle{empty}\c@page\z@
        \def\thefootnote{\fnsymbol{footnote}} }

\def\endtitlepage{\if@restonecol\twocolumn \else \newpage \fi
        \def\thefootnote{\arabic{footnote}}
        \setcounter{footnote}{0}}  

\catcode`@=12
\relax

\def\figcap{\section*{Figure Captions\markboth
        {FIGURECAPTIONS}{FIGURECAPTIONS}}\list
        {Figure \arabic{enumi}:\hfill}{\settowidth\labelwidth{Figure
999:}
        \leftmargin\labelwidth
        \advance\leftmargin\labelsep\usecounter{enumi}}}
 \relax
\def\tablecap{\section*{Table Captions\markboth
        {TABLECAPTIONS}{TABLECAPTIONS}}\list
        {Table \arabic{enumi}:\hfill}{\settowidth\labelwidth{Table
999:}
        \leftmargin\labelwidth
        \advance\leftmargin\labelsep\usecounter{enumi}}}
 \relax
\def\reflist{\section*{References\markboth
        {REFLIST}{REFLIST}}\list
        {[\arabic{enumi}]\hfill}{\settowidth\labelwidth{[999]}
        \leftmargin\labelwidth
        \advance\leftmargin\labelsep\usecounter{enumi}}}
 \relax
\makeatletter
\newcounter{pubctr}
\def\publist{\@ifnextchar[{\@publist}{\@@publist}}
\def\@publist[#1]{\list
        {[\arabic{pubctr}]\hfill}{\settowidth\labelwidth{[999]}
        \leftmargin\labelwidth
        \advance\leftmargin\labelsep
        \@nmbrlisttrue\def\@listctr{pubctr}
        \setcounter{pubctr}{#1}\addtocounter{pubctr}{-1}}}
\def\@@publist{\list
        {[\arabic{pubctr}]\hfill}{\settowidth\labelwidth{[999]}
        \leftmargin\labelwidth
        \advance\leftmargin\labelsep
        \@nmbrlisttrue\def\@listctr{pubctr}}}
 \relax
\makeatother
\newskip\humongous \humongous=0pt plus 1000pt minus 1000pt

\newif\ifdtup

\relax

\def\be{\begin{equation}}
\def\ee{\end{equation}}
\def\ba{\begin{eqnarray}}
\def\ea{\end{eqnarray}}

\def\del{\partial}

\def\a{\alpha}

\def\b{\beta}

\def\d{\delta}

\def\m{\mu}
\def\n{\nu}

\def\l{\lambda}
\def\L{\Lambda}
\def\s{\sigma}

\def\no{\noindent}

\def\qq{\qquad}

\def\IR{\relax{\text I\kern-.18em R}}

\def \ha {{1\over 2}}

\def \ov {\over}

\def\diag{{\text diag}}

\def\IR{\relax{\text I\kern-.18em R}}
\def\IL{\relax{\text I\kern-.18em L}}

\def\inv{^{\raise.15ex\hbox{${\scriptscriptstyle -}$}\kern-.05em 1}}

\begin{document}

\renewcommand{\theequation}{\thesection.\arabic{equation}}
\csname @addtoreset\endcsname{equation}{section}

\newcommand{\beq}{\begin{equation}}
\newcommand{\eeq}[1]{\label{#1}\end{equation}}
\newcommand{\ber}{\begin{eqnarray}}
\newcommand{\eer}[1]{\label{#1}\end{eqnarray}}
\newcommand{\eqn}[1]{(\ref{#1})}
\begin{titlepage}
\begin{center}

~


\vskip  .3in

{\large\bf The anisotropic $\lambda$-deformed $SU(2)$ model is
integrable}

\vskip 0.4in

{\bf Konstantinos Sfetsos}$^1$\ and\ {\bf Konstantinos Siampos}$^2$
\vskip 0.1in {\em
${}^1$Department of Nuclear and Particle Physics\\
Faculty of Physics, University of Athens,\\
Athens 15784, Greece\\
 {\tt \footnotesize ksfetsos@phys.uoa.gr}}
\vskip 0.1in

{\em ${}^2${Albert Einstein Center for Fundamental Physics,\\
Institute for Theoretical Physics, Bern University,\\
Sidlerstrasse 5, CH3012 Bern, Switzerland\\
{\tt\footnotesize siampos@itp.unibe.ch}}}

\end{center}

\vskip 0.2in

\centerline{\bf Abstract}

\no
The all-loop anisotropic Thirring model interpolates between the
 WZW model and the non-Abelian T-dual of the anisotropic
 principal chiral model.
We focus on the $SU(2)$ case and we  prove that it is classically integrable by providing its Lax pair formulation.
We derive its underlying symmetry current algebra and use it to show that the Poisson brackets of the spatial part of the
Lax pair, assume the Maillet form. In this way we procure the corresponding
$r$ and $s$ matrices which provide non-trivial solutions to the modified Yang--Baxter equation.

\newpage

\end{titlepage}

\tableofcontents


\def\baselinestretch{1.2}
\baselineskip 20 pt
\noindent


\setcounter{equation}{0}
\renewcommand{\theequation}{\thesection.\arabic{equation}}

\section{Introduction and motivation}
\label{intro}

The general class of $\s$-models whose integrability properties will be investigated was constructed in \cite{Sfetsos:2013wia}.
The corresponding action is given by
\be
S_{k,\l}(g) = S_{{\text WZW},k}(g) - {k\ov \pi} \int J^A_+ M^{-1}_{AB} J^B_-\ ,\qq M_{AB}=(\l^{-1}- D^T)_{AB}\ ,
\label{action1}
\ee
where the first term is the WZW model action for a semi-simple compact group $G$
and a group element $g\in G $ given by \cite{Witten:1983ar}
\be
S_{{\text WZW}.k}(g) = -{k\ov 2\pi} \int {\text Tr}(g^{-1} \del_+ g g^{-1} \del_- g)
+ { k\ov 6\pi} \int_B {\text Tr}(g^{-1}\mathrm{d}g)^3\ .
\label{dhwzw}
\ee
This is a CFT with two commuting current algebras at level $k$.
The second term in \eqn{action1} represents the deformation from the conformal point. Our conventions are\footnote{
The world-sheet coordinates $(\s^+,\s^-)$ and $(\tau,\sigma)$ are related by
$
\s^\pm = \tau \pm \s$, so that
$\del_0 =\del_\tau= \del_+ +\del_-$, $\del_1=\del_\s = \del_+ - \del_-
$.
}
\be
\begin{split}
J^A_{+} = {\text Tr}(t^A \del_+ g g^{-1})\,,
\quad J^A_{-} ={\text Tr}(t^A g^{-1} \del_- g )\,,\quad
D_{AB} = {\text Tr}(t_A g t_B g^{-1})\ .
\label{defsym}
\end{split}
\ee
The above action has a $\dim G$ target space with coordinates the parameters in the group element $g\in G$.
The $t_A$'s are representation matrices obeying the Lie algebra $[t_A,t_B]=f_{ABC}\, t_C$ and
normalized as ${\text Tr}(t_A t_B)=\d_{AB}$. The deviation from the WZW model is parametrized by the coupling
matrix elements $\l_{AB}$. For small such elements the Lagrangian density is proportional to the current bilinear $\l_{AB} J_+^A J_-^B$. { Hence the name $\l$-deformed models.}
The above action develops an extra local invariance under
the vector action of a subgroup $H\subset G$ when $\l_{AB}$ assumes the block
diagonal form $\l_{AB}={\text diag}(\mathbb{I}_{ab},\l_{\a\b})$, where the lower case Latin indices take
values in the Lie algebra of $H$ and the Greek ones in the coset $G/H$.
Due to this local invariance $\dim H$ degrees of freedom become redundant. Hence, $\dim H$ variables among
those parameterizing $g$ should be
gauged fixed. For vanishing $\l_{\a\b}$ the $\s$-model corresponds to the coset $G/H$ CFT.
In addition, the perturbation is driven by parafermion bilinears $\l_{\a\b} \Psi_+^\a \Psi_-^\b$,
where the $\Psi_\pm^\a$'s are gauge invariant versions of the currents $J_\pm^\a$.
The renormalization group equations for $\l_{AB}$ in the action \eqn{action1} have been computed for the isotropic case in \cite{Itsios:2014lca}
and in full generality in \cite{Sfetsos:2014jfa}. In addition, the \eqn{action1} has been used as a building block
to construct full solution of type-II supergravity in \cite{Sfetsos:2014cea} which are likely also integrable at the
string level.

\no
In this paper we are interested in investigating integrability property of the above action.
Integrability has been first proven for the isotropic case when $\l_{AB}=\l \d_{AB}$ and a general semi-simple group $G$ in
\cite{Sfetsos:2013wia}.
This was done by explicitly showing that certain
algebraic conditions developed in \cite{Balog:1993es} (based on earlier work
in \cite{Rajeev:1988hq}) were satisfied.\footnote{In this work the $\s$-model fields corresponding to
\eqn{action1} for the isotropic case and when $G=SU(2)$ were also constructed
by a brute force computation which is not generalizable in practice for larger groups.}
In addition, it has been proved that these models have an underlying Yangian symmetry \cite{Itsios:2014vfa}.
In \cite{Sfetsos:2013wia}, integrability
was also expected for the coset $SU(2)/U(1)$ case by making contact with the work of \cite{Fateev:1991bv}
where a CFT approach was utilized.
The most efficient way to prove integrability of \eqn{action1} for specific choices of the matrix $\l$ is to
{ employ} its origin via a gauging procedure as much as possible.
This was done for the { aforementioned} isotropic group case as well as for the general symmetric coset space,
for isotropic { coupling} $\l_{\a\b}=\l \d_{\a\b}$ in \cite{Hollowood:2014rla}.

In the present paper we will generalize this approach
for a symmetric matrix $\l$ and we will prove integrability for the anisotropic $SU(2)$ model for 
symmetric  matrix $\l_{AB}$.
The computation amounts to showing integrability for the diagonal matrix $\l=\diag(\l_1,\l_2,\l_3)$. In addition,
we will compute the Poisson brackets of the spatial component of the corresponding monodromy matrix and we will provide
non-trivial solutions to the modified Yang--Baxter equation.

\section{Origin of integrability}
\label{Origin}

In this section we review the construction of our models, derive the equations of motion and
set them up in such a way that investigating the existence of a Lax pair formulation becomes immediate.

\subsection{Review of the models}

We review the construction of the models by following \cite{Sfetsos:2013wia}.
The starting point is the action
\be
S(g,\tilde g) = S_{{\text WZW},k}(g)  + S_{PCM,E}(\tilde g)\ ,
\label{totact}
\ee
where the first term is the WZW action \eqn{dhwzw} and the second term is the principal chiral model (PCM) action for $G$ using a group element $\tilde g\in G$,
\be
\label{PCM.aniso}
S_{{\text PCM},E}(\tilde g) = -{1\ov \pi} \int E_{AB} {\text Tr}(t^A\tilde g^{-1} \del_+\tilde g ){\text Tr}(t^B\tilde g^{-1} \del_-\tilde g )\ .
\ee
The action \eqn{totact} is invariant under left-right current algebra symmetry of the WZW action and
a global left symmetry of the PCM.  We will gauge the same global group
\be
g\to \L^{-1} g\L \ ,\qq \tilde g\to \L^{-1} \tilde g\ ,\qq \L\in G\ .
\ee
Hence we consider the action
\be
\label{full.gauged}
S_{k,E}(g,\tilde g) = S_{{\text gWZW},k}(g,A_\pm) + S_{{\text gPCM},E}(\tilde g,A_\pm)\ ,
\ee
where
\be
\begin{split}
&S_{{\text gWZW},k}(g,A_\pm)= S_{{\text WZW},k}(g) + {k\ov \pi}\int {\text Tr}\left(A_-\del_+ g g^{-1}
- A_+ g^{-1}\del_- g\right.\\
&\phantom{xxxxxxxxxxxxxx}\left. + A_- g A_+ g^{-1} - A_-A_+\right)\ ,
\label{gtotact}
\end{split}
\ee
and
\be
S_{{\text gPCM},E}(\tilde g,A_\pm) = -{1\ov \pi}\int E_{AB} {\text Tr}(t^A \tilde g^{-1} \widetilde D_+ \tilde g)
{\text Tr}(t^B \tilde g^{-1} \widetilde D_- \tilde g)\ ,
\ee
with the covariant derivatives being
$
\widetilde D_\pm \tilde g = \del_\pm \tilde g - A_\pm \tilde g
$.
This action \eqn{gtotact} is invariant under the local symmetry
\be
\tilde g\to \L^{-1} \tilde g\ ,\qquad g\to \L^{-1} g \L\ ,
\qquad A_\pm\to \L^{-1}A_\pm \L - \L^{-1}\del_\pm \L \ .
\label{khgk2}
\ee
We will use the coupling matrix $\l$ defined as $E= k (\l^{-1}-\mathbb{I})$.
Finally we mention that the action \eqn{full.gauged} is invariant under the generalized parity symmetry
\be
\s^+\leftrightarrow \s^- \ ,\qquad g\mapsto g^{-1}\ ,\qquad \tilde g\mapsto \tilde g\ ,\qquad A_+\leftrightarrow A_-\ ,
\qq \l \mapsto \l^T\ .
\ee

\subsection{Gauge fixing and equations of motion}

We may choose the gauge $\tilde g = \mathbb{I}$. It is easily seen that the equation of
motion followed by varying $\tilde g$ is automatically satisfied.
Varying the action with respect to $A_\pm$ we find the constraints
\be
\qq D_+ g\, g^{-1} =  (\l^{-T}-\mathbb{I}) A_+ \ ,\qq g^{-1} D_- g = - (\l^{-1}-\mathbb{I}) A_-\ ,
\label{dggd}
\ee
where $D_\pm g = \del_\pm g - [A_\pm,g],$ or equivalently
\be
\label{AtoJ}
A_+ =  (\l^{-T}-D)^{-1} J_+\ ,\qq A_- =- (\l^{-1}-D^T)^{-1} J_-\ .
\ee
Varying with respect to the group element $g$ we obtain that
\be
\begin{split}
& D_-(D_+g g^{-1}) = F_{+-} \ ,\qq D_+(g^{-1}D_- g ) = F_{+-} \  ,\\
&  F_{+-} =\del_+ A_- - \del_- A_+ -[A_+,A_-]\ ,
\label{ddginv}
\end{split}
\ee
which due to
$
[D_+,D_-]g =[g,F_{+-}]
$,
turn out to be equivalent.
Substituting \eqn{dggd} into \eqn{ddginv} we obtain that
\be
D_-\left((\l^{-T}-\mathbb{I})  A_+ \right)= { F_{+-}} \ ,
\qq -D_+\left( (\l^{-1}-\mathbb{I}) A_-\right) = { F_{+-}} \ ,
\ee
which can be cast as\footnote{
In components
\begin{equation*}
P A_+=P_{BC}A_+^C\,t_B\,,\qq [P A_+,A_-]=f_{BCD}P_{CE}A_+^E A^D_-\,t_B\,,
\end{equation*}
where $P$ is an arbitrary square matrix.
}
\be
\begin{split}
\label{eomAinitial}
& \del_+ A_- - \del_- (\l^{-T} A_+) = [\l^{-T} A_+,A_-]\ ,
\\
& \del_+(\l^{-1}A_-)-\del_-A_+=[A_+,\l^{-1}A_-]\ .
\end{split}
\ee
Unless $\l= \mathbb{I}$, the $A_\pm$ are not pure gauges.
Solving for $\lambda\neq\mathbb{I}$ we obtain that
\be
\begin{split}
& \del_+ A_- =  (\mathbb{I}-\l\l^T)^{-1} \left(-\l\l^T [\l^{-T}A_+,A_-]+\l [A_+,\l^{-1}A_-]\right)\ ,
\\
&
 \del_- A_+ = (\mathbb{I}-\l^T\l)^{-1} \left(\l^T \l [A_+,\l^{-1}A_-]-\l^T [\l^{-T}A_+,A_-]\right)\ .
\label{eomA}
\end{split}
\ee
Note that for a symmetric matrix $\l$, with $\dim G$ linearly independent eigenvectors, 
it is sufficient to prove integrability using its diagonal form. To show this, we note that 
\eqn{eomAinitial} (or \eqn{eomA}) are covariant
under the orthogonal transformation 
 $\l\mapsto S\l S^T$, with $A_\pm\mapsto S A_\pm S^T$.

Our goal/effort would be to rewrite, if possible, the equations of motion \eqref{eomA}
as a Lax equation
\be
\label{Lax}
\mathrm{d}L=L\wedge L\qquad \text{or} \qquad \del_+L_--\del_-L_+=[L_+,L_-]\,,
\ee
where $L_\pm=L_\pm(\tau,\sigma,\mu)$ depend on a spectral parameter $\mu\in\mathbb{C}$.

\subsection{The current algebra}

For a gauged WZW we can define
\be
\label{Affine}
S_+  = \frac k2\left(D_+ g g^{-1} + A_+ - A_-\right)\ ,\qq S_-  = \frac k2\left(- g^{-1} D_- g + A_- - A_+\right)\ ,
\ee
which obey two commuting copies of current algebras \cite{Bowcock:1988xr,Rajeev:1988hq}
\be
\label{Bowcock}
\{S_\pm^A , S_\pm^B\} = f_{ABC} S_\pm^C \d_{\s\s'} \pm \frac k2 \d_{AB} \d'_{\s\s'}\,,\qq
\d_{\s\s'}=\d(\s-\s')\,,
\ee
where we have dropped the time dependence at usual equal time Poisson brackets.
Since the action does not depend on derivatives of $A_\pm$, its equations-of-motion
are second class constraints \cite{Hollowood:2014rla,Hollowood:2014qma}
\be
S_+ = \frac k2\left(\l^{-T} A_+ - A_-\right)\ ,\qq S_- = \frac k2\left(\l^{-1} A_- - A_+\right)\
\ee
and inversely
\be
\label{AthroughS}
\begin{split}
&A_+ =\frac2k g^{-1} \l^T (S_+ + \l S_-)\ ,\qq A_- = \frac2k \tilde g^{-1} \l (S_- + \l^T S_+)\ ,\\
&g=\mathbb{I}-\l^T\l\,,\qquad \tilde g=\mathbb{I}-\l\l^T\,,
\end{split}
\ee
where we assume that $g,\tilde g$ are positive-definite matrices. It is just a matter of algebra to rewrite the current algebras
for $S_\pm$ in the base of $A_\pm$, as we are going to present in the subsequent sections.

\section{Known integrable cases}

In this section we review the known (isotropic) integrable cases,  semi-simple group and general symmetric coset spaces, using the previous formulation.

\subsection{The isotropic group space}

As a warmup, we review the integrability for the isotropic case for a semi-simple group $G$ \cite{Sfetsos:2013wia, Hollowood:2014rla}.
Then the equations of motion for the gauge field read
\be
\del_\pm A_\mp = \pm {1\ov 1+ \l} [A_+,A_-]\ .
\ee
and a simple rescaling
\be
\label{iso.match}
A_\pm = -\ha (1+\l) I_\pm\ ,
\ee
proves the integrability. As for the Lax pair, this is given by
\be
\label{Lax.iso}
L_{\pm} = {2\ov 1+\l} {\mu\ov \mu\mp 1} A_\pm\ ,
\ee
where $\mu\in\mathbb{C}$ is the spectral parameter.

\subsubsection{Algebraic structure}

Employing \eqref{Bowcock}, \eqref{AthroughS} and \eqref{iso.match},
we find the Poisson brackets for $I_\pm$ \cite{Balog:1993es}
\be
\label{algebra.iso}
\begin{split}
&\{I_\pm^A,I_\pm^B\}=e^2\,f_{ABC}\left(I_\mp^C-(1+2x)I^C_\pm\right)\d_{12}\pm\,2e^2\d_{AB}\d'_{12}\ ,
\\
&\{I_\pm^A,I_\mp^B\}=-e^2\,f_{ABC}\left(I_+^C+I_-^C\right)\,\d_{12}\ ,
\end{split}
\ee
where
\be
\label{algebra.iso.defs}
e=\frac{2\lambda}{\sqrt{k(1-\lambda^2)}(1+\lambda)}\,,\qquad x=\frac{1+\lambda^2}{2\lambda}>1\ ,
\ee
{  where the deformation parameter is a root of unity \cite{Hollowood:2014rla}.}
We note that the same underlying structure, but with $-1<x<1$, corresponds to integrable deformations of the $\s$-model \cite{Klimcik:2008eq} constructed in \cite{Delduc:2013fga,Delduc:2014uaa}, where the  deformation parameter is {  real}.
The corresponding quantum properties at one-loop were studied in \cite{Squellari:2014jfa,Arutyunov:2013ega,Hoare:2014pna}.

There are two interesting limits.
Expanding $\l$ near zero and rescaling $I_\pm^A\mapsto-2e^2x I_\pm^A$ we find that
\be
\label{iso.conformal}
\{I^A_\pm,I^B_\pm\}=f_{ABC}\,I^C_\pm\,\d_{\s\s'}\pm \frac k2\,\d_{AB}\d'_{\s\s'}\,,\quad \{I_+^A,I_-^B\}=0\,.
\ee
These are two commuting current algebras in accordance with the fact that in this limit the $\s$-model
corresponds to a CFT.

Parameterizing $\l$ as $ \l=k(k+ \varepsilon)^{-1}$ and then letting $k\gg1$, we find the algebra of the non-Abelian T-dual of the PCM on $G$
\be
\label{iso.PCM}
\begin{split}
&\{I_\pm^A,I_\pm^B\}=\frac{1}{2\varepsilon}f_{ABC}(I_\mp^C-3I_\pm^C)\d_{12}\pm\frac1\varepsilon\d_{AB}\d'_{12}\,,\\
&\{I_\pm^A,I_\mp^B\}=-\frac{1}{2\varepsilon}f_{ABC}(I_+^C+I_-^C)\d_{12}\ .
\end{split}
\ee
This is the same as the algebra for the PCM for $G$, in accordance with the fact that the two cases
are related by a canonical transformation.

\subsection{The isotropic symmetric coset}
\label{int.symmetric}

Let us consider a semi-simple group $G$ and its decomposition to a semi-simple subgroup $H$ and a
symmetric coset $G/H$.\footnote{In the conventions of section \ref{intro} we
denote subgroup indices by Latin letters and coset indices by Greek letters.} Take the case
where the matrix $\lambda$ has elements
$\lambda_{ab}=\lambda_H\,\delta_{ab}\,, \lambda_{\alpha\beta}=\lambda_{G/H}\,\delta_{\alpha\beta}\,.$
The restriction to symmetric cosets translates to structure constants $f_{\alpha\beta\gamma}=0,$
whereas $f_{abc},f_{\alpha\beta c}\neq0.$ For $\lambda_H,\lambda_{G/H}\neq1$ we have that \eqref{eomAinitial} or
\eqref{eomA} read\footnote{Note that for $\lambda_{G/H}=1,$ $\l_H$ turns to be one for finiteness
of the expressions. For general cosets $G/H$ the equations for $B_\pm$
contain the additional term $\pm(1+\l_{G/H})^{-1}[B_+,B_-]$.}
\be
\label{eom.decompose}
\begin{split}
&\del_\pm A_\mp=\pm (1+\lambda_H)^{-1}\left([A_+,A_-]+\frac{\l_H}{\l_{G/H}}[B_+,B_-]\right)\,,\\
&\del_\pm B_\mp=\frac{1}{\l_H(1-\lambda_{G/H}^2)}\left((\l_{G/H}^2-\l_H)[B_\mp,A_\pm]+\l_{G/H}(1-\l_H)[B_\pm,A_\mp]\right)\,,
\end{split}
\ee
where $A_\pm$ and $B_\pm$ are Lie algebra valued one forms ($A_\pm=A^a_\pm t_a, B_\pm=B^\a_\pm t_\a$), on the subgroup and coset respectively.\footnote{
We have tried to construct a  Lax pair for  \eqref{eom.decompose}
of the form $L_\pm = a_\pm A_\pm + b_\pm B_\pm$ where the coefficients are constants. For $\l_H\neq 1$ and for non-Abelian
subgroup $H$ one obtains a linear algebraic inhomogeneous system with has a unique solution. This implies that within this
ansatz for the Lax pair one cannot prove integrability.}
The above consideration is drastically modified in the two cases we have excluded.
The first special case is when $\lambda_H=1$. In this singular
limit we have to use \eqref{eomAinitial} and the equations of motion simplify drastically
\be
\label{eom.symmetric}
\begin{split}
&\del_+ A_--\del_-A_+=[A_+,A_-]+\frac{1}{\l_{G/H}}[B_+,B_-]\,,\\
&\del_\pm B_\mp=-[B_\mp,A_\pm]\,,
\end{split}
\ee
and the two eom for $A_\pm$ in \eqref{eom.decompose} are replaced by their difference. This case
was shown to be integrable in \cite{Hollowood:2014rla} with Lax pair given by
\be
L_\pm=A_\pm+\frac{\mu^{\pm1}}{\sqrt{\l_{G/H}}}\,B_\pm\ ,
\ee
where $\mu\in\mathbb{C}$. It can be readily checked that then \eqn{Lax} is satisfied.

\section{The anisotropic $SU(2)$ case}

In this section we consider the other special case in which the subgroup $H$ is Abelian. In addition to demanding that
the space $G/H$ is symmetric, restrict our considerations to the group case $SU(2)$. We will consider the cases of a 
symmetric matrix $\l_{AB}$, $A,B=1,2,3$. Then as explained before it is sufficient to consider the case with
\be
\label{diagonal.su2}
\l =\diag (\l_1,\l_2,\l_3)\ ,
\ee
that is the fully anisotropic albeit diagonal case.
In this case the generators are $t_A=-i\,\s_A/\sqrt{2},$ where $\s_A$ are the Pauli matrices, so that $f_{ABC}=\sqrt{2}\, \varepsilon_{ABC}$.
Then a straightforward computation shows that
\be
\label{eomsu2}
\del_\pm A_\mp^1 = {\sqrt{2}\,\l_1\ov (1-\l_1^2)\l_2 \l_3} \left[(\l_2-\l_1\l_3)A_\pm^2 A_\mp^3 -
(\l_3-\l_1\l_2) A_\pm^3 A_\mp^2\right]\ ,
\ee
and cyclic in $1,2$ and $3$.

This is in agreement with \eqref{eom.decompose}, for $\l_H=\l_1,\l_{G/H}=\l_2=\l_3,$
where $H=U(1)$ and $SU(2)/U(1)$ symmetric coset. Moreover, along the results of the section \ref{int.symmetric} and \cite{Hollowood:2014rla},
the coset limit $\l_1=1$ is integrable and for compatibility $\l_2=\l_3$.
As shown in \cite{Sfetsos:2013wia} expanding the $\l_i$'s around one, we get
the non-Abelian T-dual of the anisotropic PCM for $SU(2)$, which is integrable
due to the fact that the PCM  is integrable \cite{Cherednik:1981df,hlavaty} and non-Abelian T-duality preserves integrability
\cite{Mohammedi:2008vd}.\footnote{{ We provide a detailed self-contained proof for the general case in Appendix \ref{PCM.non.abelian}.}}
All these are signals that the more general case we consider here with \eqn{diagonal.su2} is likely integrable as well.

Let's define a  convenient set of fields given by
\be
\label{su2.rescale}
X_\pm^1 = {A^1_\pm \ov \l_1 \sqrt{(1-\l_2^2)(1-\l_3^2)} }\
\ee
and cyclic in $1,2$ and $3$. Then
we assume the following expression for a Lax pair
\be
\label{Lax.su2}
L^A_\pm(\tau,\sigma;\mu)=v_\pm^A(\mu)\,X^A_\pm\,,
\ee
where $\mu\in\mathbb{C}$ and $v_\pm^A$ satisfy six non-linear equations
\be
\label{system}
c_2v^1_\mp+c_3v_\pm^1=v_\pm^2v_\mp^3\,,\quad
c_3v^2_\mp+c_1v_\pm^2=v_\pm^3v_\mp^1\,,\quad
c_1v^3_\mp+c_2v_\pm^3=v_\pm^1v_\mp^2\ ,
\ee
with $c_1=\l_1-\l_2\l_3$ and cyclic in $1,2$ and $3$.
This system turns out to
have a one parameter solution.  To prove this, we solve for example
the first and the fourth with respect to $v^{1,2}_-$
and we do the same by solving the fifth and the sixth.
By equating these alternative expressions  for $v^{1,2}_-$
we find that
$
(v_+^1)^2-(v_+^2)^2=c_1^2-c_2^2
$.
Working analogously we find two more conditions following by cyclic permutation of $1,2$ and $3$
and three analogue expressions for $v_-^A$ through a parity transformation $v_+^A\mapsto v_-^A$. Hence all together we have the
conditions
\be
(v_\pm^1)^2-(v_\pm^2)^2=c_1^2-c_2^2\ ,\quad
(v_\pm^2)^2-(v_\pm^3)^2=c_2^2-c_3^2\ , \quad (v_\pm^3)^2-(v_\pm^1)^2=c_3^2-c_1^2\ .
\ee
We proceed by solving them as
\be
\label{co4}
v^A_\pm = \sqrt{z_\pm + c_A^2}\  ,\qq z_\pm \in \mathbb{C}\ ,\qq A=1,2,3\ .
\ee
Plugging the latter in \eqref{system} and after some algebraic manipulations, we find one more
independent condition for $z_\pm$
\be
(z_+z_- - c_1^2 c_2^2 - c_2^2 c_3^2 - c_3^2 c_1^2)^2 = 4 c_1^2  c_2^2 c_3^2(z_+ +z_- + c_1^2+c_2^2 +c_3^2)\ .
\ee
This condition determines $z_+$ in terms of an arbitrary complex number $z_-$ or vise versa and so we have proved
that there is a spectral parameter ($z_+$ or $z_-$).
As a check, in the isotropic case, where $\l_A=\l$, using \eqref{su2.rescale} we find that the construction yields \eqref{Lax.iso}
\be
v_\pm=2c\frac{\mu}{\mu\mp1}\,,\qq z_\pm=c^2\frac{(3\mu\mp1)(\mu\pm1)}{(\mu\mp1)^2}\,,\qq c=\l(1-\l)\,,\qq \mu\in\mathbb{C}\,.
\ee

\subsection{The Poisson algebra}

Employing \eqref{Bowcock} and \eqref{AthroughS}
we find the Poisson brackets for the currents
\be
\begin{split}
\label{su2}
&\{A_\pm^1,A_\mp^2\}=\frac{2\sqrt{2}\l_1\l_2}{k(1-\l_1^2)(1-\l_2^2)\l_3}
\left(c_2 A_\pm^3 + c_1 A_\mp^3 \right)\d_{12}\ ,
\\
&\{A_\pm^1,A_\pm^1\}=\pm\frac{2\l_1^2}{k(1-\l_1^2)}\,\d'_{12}\,,\\
&\{A^1_\pm,A^2_\pm\}=-\frac{2\sqrt{2}\l_1\l_2}{k(1-\l_1^2)(1-\l_2^2)\l_3}
\left(c_3 A_\mp^3 - (1-\l_1\l_2\l_3) A_\pm^3 \right)\,\d_{12}\ ,
\end{split}
\ee
and with cyclic permutations in $1,2$ and $3$ for the other pairs. For consistency
we have checked that they satisfy the Jacobi identity.

Rescaling the gauge fields $A_\pm^A\mapsto\l_A A_\pm^A,$ we can easily take the limit $\l_A\to0$
\be
\label{su2.conformal}
\{A_\pm^1,A_\pm^2\}=\frac{2\sqrt{2}}{k}A_\pm^3\d_{12}\,,\qquad \{A_\pm^1,A_\pm^1\}=\pm\frac{2}{k}\d'_{12}\,,\qquad
 \{A_\pm^1,A_\mp^2\}=0\,,
\ee
and with cyclic permutations in $1,2$ and $3$ for the other pairs.
These expressions can be also obtained from \eqref{iso.conformal} by an appropriate rescaling.

Expanding $\l_A$ near the identity, we find the algebra of the non-Abelian T-dual of the PCM on $SU(2)$
\be
\label{su2.PCM}
\begin{split}
&\{A_\pm^1,A_\mp^2\}=\frac{1}{\sqrt{2}\varepsilon_1\varepsilon_2}
\left(A^3_\pm(\varepsilon_3+\varepsilon_1-\varepsilon_2)+A_\mp^3(\varepsilon_2+\varepsilon_3-\varepsilon_1)\right)\d_{12}\,,\\
&\{A_\pm^1,A^1_\pm\}=\pm\frac{\delta'_{12}}{\varepsilon_1}\,,\\
&\{A^1_\pm,A^2_\pm\}=-\frac{1}{\sqrt{2}\varepsilon_1\varepsilon_2}
\left(A_\mp^3(\varepsilon_1+\varepsilon_2-\varepsilon_3)-A_\pm^3(\varepsilon_1+\varepsilon_2+\varepsilon_3)\right)\d_{12}\,,
\end{split}
\ee
where we have let $\displaystyle \l_A=1-{\varepsilon_A\ov k}$, for $k\gg1$. This algebra should be equivalent
to the anisotropic PCM since they are related by a canonical transformation.

In the isotropic case, where $\varepsilon_A=\varepsilon$, it is in accordance with \eqref{iso.PCM} under the identification given in \eqref{iso.match}.

\subsection{Maillet brackets}

Following Sklyanin \cite{Evgeny}, we compute the equal time Poisson bracket of $L_1$:
\be
\label{Maillet.lhs}
\{ L^{(1)}_1 (\sigma_1;\mu), L^{(2)}_1 (\sigma_2;\nu) \} \, = \, \{ L_1^B (\sigma_1;\mu), L_1^C (\sigma_2;\nu) \} \, t_B \otimes t_C\,,
\ee
where $L_1=L_1^A\,t_A$ and the superscript in parenthesis denotes the vector spaces on which the matrices act.\footnote{
In brief:
\begin{equation*}
\begin{split}
&M^{(1)}=M\otimes\mathbb{I}\,,\qquad M^{(2)}=\mathbb{I}\otimes M\,,\qquad M=M_A t_A\,,\\
&m^{(12)} \, = \, m_{AB} \, t_A \otimes t_B \otimes \mathbb{I}, \quad  m^{(13)} \, = \, m_{AB} \, t_A \otimes \mathbb{I} \otimes t_B, \quad m^{(23)} \, = \, m_{AB} \, \mathbb{I} \otimes t_A \otimes t_B,\nonumber
\end{split}
\end{equation*}
for an arbitrary matrix $m=m_{AB}\,t_A\otimes t_B.$
}
These brackets assume the {\it Maillet}
form \cite{Maillet}
\be
\label{Maillet.rhs}
\left([r_{-\mu\nu}, L^{(1)}_{1}(\sigma_1;\mu)]
+[r_{+\mu\nu},  L^{(2)}_{1}(\sigma_1;\nu)]\right)\delta_{12} -2s_{\mu\nu}\,\delta'_{12}\,,
\ee
where  $r_{\pm\mu\nu}=r_{\mu\nu}\pm s_{\mu\nu}$ and $r_{\mu\nu}, s_{\mu\nu}$ are matrices on the basis $t_B\otimes t_C$ depending on $(\mu,\nu)$.
This is guaranteed to give a consistent Poisson structure, provided the Jacobi identities for these brackets are obeyed. This enforces $r_{\pm\mu\nu}$ to satisfy the modified classical Yang--Baxter relation
\be
\begin{split}
\label{mcYBE}
[r^{(13)}_{+\nu_1\nu_3}, r^{(12)}_{-\nu_1\nu_2}] + [r^{(23)}_{+\nu_2\nu_3}, r^{(12)}_{+\nu_1\nu_2}] +
[r^{(23)}_{+\nu_2\nu_3}, r^{(13)}_{+\nu_1\nu_3}] =0\ .
\end{split}
\ee
The non-vanishing coefficient of the $\d'$ term in \eqn{Maillet.rhs} is responsible for the above modification,
appearance of $s_{\mu\nu}$, of the classical Yang--Baxter relation.
In what follows within this section, we shall rewrite \eqref{Maillet.lhs} and \eqref{Maillet.rhs} and retrieve 
$r_{\pm\mu\nu}, s_{\mu\nu}$.

Expanding the Poisson bracket \eqref{Maillet.lhs} we find that
\be
\label{Maillet.lhs1}
v^B_{+\mu}v^C_{+\nu}\{X_+^B,X_+^C\}+v^B_{-\mu}v^C_{-\nu}\{X_-^B,X_-^C\}
-v^B_{+\mu}v^C_{-\nu}\{X_+^B,X_-^C\}-v^B_{-\mu}v^C_{+\nu}\{X_-^B,X_+^C\}\,.
\ee
As noted this will take the form of \eqref{Maillet.rhs}.
To proceed we decompose this in two terms corresponding to $\delta'_{12}$ and $\delta_{12}$.

To compute the coefficient of $\delta_{12}'$ we use \eqref{su2.PCM} with \eqn{su2.rescale} and \eqref{Maillet.lhs1}.
We find that that $s_{\mu\nu}$ has only diagonal elements
\be
s^{11}_{\mu\nu}=-\frac{1}{k(1-\l_1^2)(1-\l_2^2)(1-\l_3^2)}\left(v_{+\mu}^1v_{+\nu}^1
-v_{-\mu}^1v_{-\nu}^1\right)\,,
\ee
and with cyclic permutations in $1,2$ and $3$ the other two. Note that they are symmetric under
the exchange of $\mu,\nu$ as expected by the antisymmetry of the Poisson bracket \cite{Maillet}.

To compute the coefficient of $\delta_{12}$ we expand in the $t_B\otimes t_C$ basis and we obtain
\be
\label{Maillet.rhs1}
\del_1r_{-\mu\nu}^{BC}+\sqrt{2}\,\varepsilon_{ABD}\, r_{-\mu\nu}^{DC}L_{1\mu}^A-\sqrt{2}\,\varepsilon_{ADC}\, r_{+\mu\nu}^{BD}L_{1\nu}^A\,.
\ee
Using \eqref{Maillet.lhs1} and \eqref{Maillet.rhs1}, we find that $r_{\mu\nu}$ has only diagonal elements.
Analyzing the $t_1\otimes t_2$ component and  heavily using \eqref{system} we find that
\be
\begin{split}
&\frac{k}{2}(1-\l_1^2)(1-\l_2^2)(1-\l_3^2)\left(v_{+\mu}^3v_{-\nu}^3-v_{+\nu}^3v_{-\mu}^3\right)
r_{+\mu\nu}^{11}=\\
&(c_3(1-\l_1\l_2\l_3)-c_1c_2)(v_{+\mu}^2v_{+\nu}^2+v_{-\mu}^2v_{-\nu}^2)+\\
&(c_1(1-\l_1\l_2\l_3)-c_2c_3)(v_{+\mu}^2v_{-\nu}^2+v_{-\mu}^2v_{+\nu}^2)\\
&-c_1\left(v_{+\mu}^1v_{-\nu}^2v_{+\mu}^3+v_{-\mu}^1v_{+\nu}^2v_{-\mu}^3\right)
-c_3\left(v_{+\mu}^1v_{+\nu}^2v_{+\mu}^3+v_{-\mu}^1v_{-\nu}^2v_{-\mu}^3\right)\,,
\end{split}
\ee
and
\be
\begin{split}
&\frac k2(1-\l_1^2)(1-\l_2^2)(1-\l_3^2)\left(v_{+\mu}^3v_{-\nu}^3-v_{+\nu}^3v_{-\mu}^3\right)
r_{-\mu\nu}^{22}=\\
&(c_3(1-\l_1\l_2\l_3)-c_1c_2)(v_{+\mu}^1v_{+\nu}^1+v_{-\mu}^1v_{-\nu}^1)+\\
&(c_2(1-\l_1\l_2\l_3)-c_1c_3)(v_{+\mu}^1v_{-\nu}^1+v_{-\mu}^1v_{+\nu}^1)\\
&-c_2\left(v_{+\mu}^1v_{-\nu}^2v_{-\nu}^3+v_{-\mu}^1v_{+\nu}^2v_{+\nu}^3\right)
-c_3\left(v_{+\mu}^1v_{+\nu}^2v_{+\nu}^3+v_{-\mu}^1v_{-\nu}^2v_{-\nu}^3\right)\ ,
\end{split}
\ee
which expressions determine $r^{11}_{+\m\n}$ and $r^{22}_{-\m\n}$. The rest of the coefficients
are determined by a cyclic permutations in $1,2$ and $3$. Although cyclicity is not profound
in the above expressions, we
can restore it by adding the corresponding equivalent expressions evaluated by the other components.

Finally, as it was stated in \eqref{mcYBE},  $r_{\pm\mu\nu}$ satisfy the modified classical Yang--Baxter equation, which in
our case reduces to six equations given compactly by
\be
r_{+\nu_1\nu_2}^{AA}r_{+\nu_2\nu_3}^{CC}
=r_{-\nu_1\nu_2}^{BB}r_{+\nu_1\nu_3}^{CC}+r_{+\nu_1\nu_3}^{AA}r_{+\nu_2\nu_3}^{BB}\,,\qquad
A\neq B\neq C\ .
\ee
The explicit form of the equations can be extracted from the coefficients of the combination $\varepsilon_{ABC}\,t_A\otimes t_B\otimes t_C$.
We have checked that this condition is indeed satisfied through a heavy use of \eqref{system}.

\section{Conclusion and outlook}
\label{Conclusions}

In this paper we proved that the $\s$-model action \eqref{action1} for the group $SU(2)$ and
for a symmetric coupling matrix $\l_{AB}$ is classically integrable. We achieved this by explicitly constructing
the spectral depending Lax pair \eqref{Lax.su2} and thus giving rise to an infinite number of conserved charges.
We computed the Poisson bracket of the spatial part $L_1$ of the Lax pair and demonstrated that it assumes the 
{ {\it Maillet-type} form \cite{Maillet,Maillet2}} from which we read off the $r$ and $s$ matrices satisfying the modified Yang--Baxter equation, arising from the Jacobi identity for these Poisson brackets.
Our result establish an integrable interpolation between the WZW model (CFT) and the non-Abelian T-dual for the anisotropic PCM for $SU(2)$.

In the context of $\l$-deformations, integrability has been proven so far for three cases:
The isotropic case, i.e. single coupling and any group $G$, the symmetric coset case $G/H$ again for a single coupling,
and finally for the anisotropic $SU(2)$ case with a symmetric coupling matrix in the present paper.
The latter case is special as it possesses only Abelian subgroups which seems to be at the root of the integrability proof we have achieved. One may wonder
if there exist other cases, based either on groups or on (non) symmetric cosets for which specific choices of the matrix $\l$ may render the corresponding $\s$-model as classically integrable. A starting point in this direction could be to examine if with
the right amount of torsion non-symmetric coset spaces may prove integrable.
In fact the $U(3)/U(1)^3$ non-symmetric coset was recently shown to belong in this category \cite{Bykov:2014efa}, 
{although the two-form takes imaginary values.}

\section*{Acknowledgements}

The research of K. Sfetsos is implemented
under the \textsl{ARISTEIA} action (D.654 of GGET) of the \textsl{operational
programme education and lifelong learning} and is co-funded by the
European Social Fund (ESF) and National Resources (2007-2013).
The research of K. Siampos has been supported by the Swiss National Science Foundation.
K.~Siampos also acknowledges the \textsl{Germaine de Stael}
France--Swiss bilateral program for financial support.
The authors thank the University of Patras for kind hospitality where this work was initiated.

\appendix

\section{Anisotropic PCM and its non-abelian T-dual}
\label{PCM.non.abelian}

In this appendix we prove that the equations of motion and the Bianchi identities 
of the anisotropic PCM and its non-abelian T-dual are mapped to each other.

The anisotropic PCM action \eqref{PCM.aniso} can be reformulated as
\be
S = \frac{1}{2\pi} \int {\text Tr}\left( j\wedge \star G j -  j \wedge B j \right)\,,\quad j=g^{-1}\mathrm{d}g\,, \quad
E = G + B\,.
\label{ggkkli1}
\ee
Varying with respect to $g$ we find the eom
\be
\label{PCM.eom}
G\, \mathrm{d}\star j = B\,  \mathrm{d} j  -(G\star j- B j)\wedge j - j\wedge (G\star j- B j)\,,
\ee
plus the flatness condition for $j$
\be
\label{PCM.flat}
\mathrm{d} j + j \wedge j = 0 \ .
\ee

We would like to show that these follow from \eqn{eomA} by letting $k\gg1$
\be
\l = \mathbb{I} - {E\ov k} + {\cal O}\left(1\ov k^2\right) \,,
\ee
and keeping the leading term in the $\displaystyle {1\ov k}$ expansion.
Indeed one easily obtains that
\be
\begin{split}
& \del_+ A_- = (E+E^T)^{-1} \left( E^T[A_+,A_-] + [A_+,E A_-] - [E^T A_+,A_-] \right) \ ,\\
& \del_- A_+ = (E+E^T)^{-1} \left( -E[A_+,A_-] + [A_+,E A_-] - [E^T A_+,A_-] \right)\ .
\end{split}
\ee
These can be rewritten as
\be
\del_+ A_- - \del_- A_+ =[A_+,A_-] \quad  \text{or} \quad \mathrm{d}A = A\wedge A \
\ee
and
\be
E\,\partial_+A_-+E^T\partial_-A_+=[A_+,EA_-]-[E^TA_+,A_-]\,.
\ee
It is elementary to prove that these can be mapped to \eqref{PCM.eom}, \eqref{PCM.flat} for $A= -j$.

\end{document}